\documentclass[twocolumn,superscriptaddress,showpacs,10pt,prl]{revtex4}
\usepackage{graphicx}%
\usepackage{dcolumn}
\usepackage{amsmath}
\usepackage{ifthen}

\newboolean{camera}   % Boolean variable, camera ready version: true
\setboolean{camera}{true}\makeatletter\appdef\set@pica@hook{\textheight = 22.9cm}\makeatother

\DeclareGraphicsExtensions{.eps}
\ifthenelse{\boolean{camera}}{          % for camera
\setlength{\textwidth}{17cm}
\setlength{\columnsep}{1cm}
\addtolength{\topmargin}{-.5cm}
\addtolength{\headsep}{1.5cm}
\makeatletter
\def\frontmatter@abstractwidth{\textwidth}
\makeatother}
{} % for
\makeatletter
\def\@pacs@name{Keywords: }
\makeatother

\newcommand{\re}{\text{Re}}
\newcommand{\im}{\text{Im}}

\begin{document}

\ifthenelse{\boolean{camera}}{
  \title{\vspace*{1cm}
  High Frequency Conductivity in the Quantum Hall Effect \vspace*{-1em}}}
{\title{High Frequency Conductivity in the Quantum Hall Effect}}

\author{F.~Hohls}
\ifthenelse{\boolean{camera}}{\email[e-mail: ]{hohls@nano.uni-hannover.de}}{}
\author{U.~Zeitler}%
\author{R.~J.~Haug}
\affiliation{%
Institut f. Festk\"orperphysik,
Universit\"at Hannover, Appelstr. 2, 30167 Hannover, Germany
}%

\author{K.~Pierz}
\affiliation{
Physikalisch-Technische Bundesanstalt, Bundesallee 100, 
38116 Braunschweig, Germany 
}%

\ifthenelse{\boolean{camera}}{}{\date{\today}} % Datum für preprint

\newcommand{\mykeywords}{quantum Hall effect, high frequency conductivity,
  scaling theory}     % Definition of keywords

\begin{abstract}
We present high frequency measurements of the diagonal conductivity 
$\sigma_{xx}$
of a two dimensional electron system in the integer quantum Hall regime. 
The width
of the $\sigma_{xx}$ peaks between QHE minima is analyzed within the
framework of scaling theory using both 
temperature ($T=100-700$~mK) and frequency ($f \leq 6$~GHz) in a two
parameter scaling ansatz. 
For the plateau transition width $\Delta\nu$ we find scaling behaviour
for both its temperature dependence as well as its frequency dependence.
However, the corresponding scaling exponent for temperature 
($\kappa=0.42$)
significantly differs from the one deduced for frequency scaling ($c=0.6$).
Additionally we use the high frequency experiments to suppress the contact
resistances that strongly influences DC measurements. 
We find an intrinsic critical
conductivity $\sigma_c=0.17\,e^2/h$, virtually independent of temperature
and filling factor, and deviating significantly from the proposed universal
value $0.5\,e^2/h$.

\ifthenelse{\boolean{camera}}{}   % takes care of keywords and corresponding
{ \vspace{1em} Keywords: \mykeywords\\[1em] % author adress for  preprint
  Corresponding author:\\ 
  Frank Hohls\\Universitaet Hannover, Institut f. Festkoerperphysik\\
  Appelstr.2, 30167 Hannover, Germany\\
  e-mail: hohls@nano.uni-hannover.de, Fax.: ++49 511 762 2904 }
\end{abstract}

\pacs{\mykeywords}  % pacs plays keyword

\maketitle

\newlength{\plotwidth}          % variable with maximum width for figures
\setlength{\plotwidth}{0.7\linewidth} % makes figures smaller for preprint
\ifthenelse{\boolean{camera}}{\thispagestyle{myheadings}% page number on first page
   \setlength{\plotwidth}{\linewidth}}{} % for camera figures have linewidth

%\mysection{Introduction}
%
Fundamental progress in the understanding of the 
properties of a two-dimensional
electron system (2DES) in high magnetic fields was brought by the 
application of scaling theory to the transition between quantum Hall
plateaus (review \cite{huckestein95} and references therein). 
Within this theoretical picture 
the step in the Hall conductivity $\sigma_{xy}$ 
between quantized values
and the corresponding maximum of the longitudinal conductivity 
$\sigma_{xx}$ are governed by a diverging
localization length $\xi\propto |\nu-\nu_c|^{-\gamma}$ with filling factor
$\nu$ and critical point $\nu_c$ near half integer filling. 
The exponent $\gamma$ is believed to be universal with a value of 
$\gamma = 2.3$ found in numerical studies and size
scaling experiments. 
The conductivities are given by scaling functions 
$\sigma_{\alpha\beta}\left(L_{\text{eff}}/\xi\right)$.
$L_{\text{eff}}$ is an effective length scale governed by sample size, 
temperature or frequency. The natural temperature dependent length scale
is determined by the phase coherence length $L_\Phi\propto T^{-p/2}$
with $p=2$ deduced from experiment.
Applying a high frequency to the system introduces an additional length
scale, the dynamic length $L_f\propto f^{-1/z}$ with dynamic exponent
$z=1$ found from numerical studies \cite{huckestein99}.
As a consequence of the length scales defined above the width $\Delta \nu$
of the conductivity peak around $\nu_c$ is predicted to follow power laws,
$\Delta \nu(T)\propto T^\kappa$ with $\kappa = p/2\gamma$ and 
$\Delta \nu(f) \propto f^c$ with $c=1/z\gamma$.
The temperature 
dependence has been subject to many experiments, mostly with the result
of $\kappa\approx0.43$ from which follows $p=2$.
Only few experiments
\cite{engel93,balaban98,kuchar00} addressed the frequency dependence and
yielded contradicting results: While Engel \textsl{et al.} \cite{engel93}
measured scaling behaviour with $c\approx 0.43$ consistent with $z=1$, 
an experiment of Balaban \textsl{et al.} \cite{balaban98} contradicts scaling.

Right at the critical point $\sigma_{xx}$ has its maximum value and 
is named critical conductivity $\sigma_c$.
Following scaling theory this value should be independent of temperature,
frequency and
of the viewed transition identified by $\nu_c$. Even further there are
analytical arguments and numerical calculations 
for a sample independent universal critical conductivity
value of $\sigma_c=0.5\,e^2/h$ (references in \cite{huckestein95}),  
but most experiments do not even follow the first prediction.

In this paper we report frequency and temperature
dependent measurements with $f=100\,\text{kHz}-6\,\text{GHz}$ and 
$T=100-700$~mK 
which are analyzed within the framework of scaling theory.
The measured transition widths follow scaling behaviour and are analyzed
using a two parameter
scaling ansatz. We find different exponents $\kappa=0.42\pm0.05$ 
and $c=0.6\pm0.1$ for frequency and temperature dependence. 

The critical conductivities 
$\sigma_c$ at low frequencies $f<1$~GHz are temperature, frequency and
transition dependent which can be understood by contact effects. At high
frequency contact effects are negligible and we measure a transition
independent non-universal value $\sigma_c=0.17\,e^2/h$. 

\begin{figure}[tb]
  \resizebox{\plotwidth}{!}{\rotatebox{0}{\includegraphics{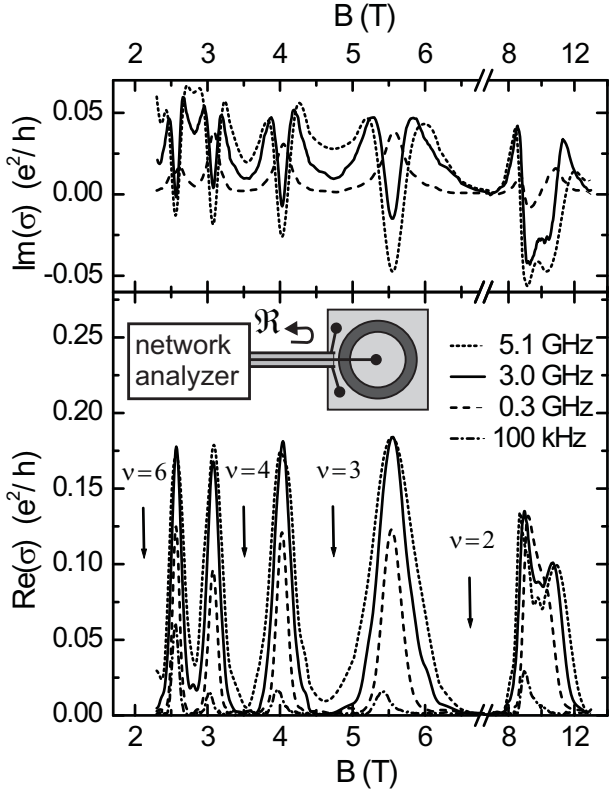}}}
  \caption{Real and imaginary part of conductivity for different frequencies.}
  \label{data}
\end{figure}

%\mysection{Experiment}
%
The sample used in the present work is an \mbox{AlGaAs/GaAs} heterostructure
grown by molecular-beam epitaxy containing 
a 2DES with a moderate electron mobility 
$\mu_e = 35$~m$^2$/Vs and an electron density $n_e=3.3\cdot 10^{15}$~m$^{-2}$. 
The sample was
patterned in a Corbino geometry with ohmic contacts fabricated by standard
Ni/Au/Ge alloy annealing (see inset in Fig. 
\ref{data}).  The Corbino
geometry allows a two-point measurement of the diagonal conductivity
$\sigma_{xx}$ which will be referred to as $\sigma$ throughout this paper. 
The two-point measurement
is the only possible type of measurement at microwave frequencies. 
The relation between the conductance $G=I/U$ with current $I$
and voltage $U$ and the conductivity $\sigma$ is given by 
$\sigma = \frac{G}{2\pi}\ln(\frac{r_2}{r_1})$
where $r_2=820$~$\mu$m is the outer radius of the Corbino ring
and $r_1=800$~$\mu$m is its inner radius.

The sample was used as termination of a standard coaxial line with impedance
$Z_0=50\,\Omega$. 
The deviation of the sample impedance $Z=1/G$ from $Z_0$
leads to reflection of a microwave signal fed into the line. The reflection
coefficient at the end of the line is given by 
$\mathcal{R}_p = \frac{Z-Z_0}{Z+Z_0}$. By measuring this reflection coefficient
we are able to deduce the conductivity $\sigma$. 
The setup would be most sensitive to changes in $\sigma$ if the sample 
impedance $Z=1/G$ was close to the line impedance $Z_0=50\,\Omega$. This is
the reason why we chose such extreme aspect ratio of the Corbino disc
yielding $Z\gtrsim 1\;\text{k}\Omega$ 
for the expected conductivity $\sigma \lesssim 0.5 e^2/h$. We are still far
away from $Z=Z_0$, but a smaller ring width
would lead to size effects.
Sample and coaxial line were fitted into a dilution refrigerator with base
temperature $T_S < 50$~mK. 
Great care was taken on the thermal sinking of the coaxial line. 
An important point is a careful characterization of the frequency dependent 
losses, phase shifts and connector reflections of the coaxial line. With this 
information we are able to extract the frequency dependent sample reflection
coefficient $\mathcal{R}_p$ from the total reflection $\mathcal{R}$ of the
line measured with a network analyzer with frequency range $100$~kHz to 
$6$~GHz. 
The result of such a measurement of the sample conductivity as a function 
of the magnetic field is shown in figure \ref{data}. Our measurement 
technique naturally gives access to real and imaginary part of $\sigma$.

%\mysection{Amplitudes}
%
The measured amplitude of the 
Shubnikov-de Haas (SdH) oscillations of the real part of the conductivity 
$\re(\sigma)$ strongly
rises from 100~kHz to 300~MHz (figure 2). 
The magnetic field dependence of the imaginary part $\im(\sigma)$
in this low frequency 
range is also distinct from the high frequency behaviour and shows 
SdH oscillation in phase with the real part. 
This contradicts scaling theory where frequency effects should  
be negligible for $hf \ll k_BT$. With an electron temperature
$T_e \geq 100$~mK a marginal frequency dependence for $f < 1$~GHz would
follow.
\begin{figure}[tb]
  \begin{center}
  \resizebox{0.9\plotwidth}{!}{\rotatebox{270}{\includegraphics{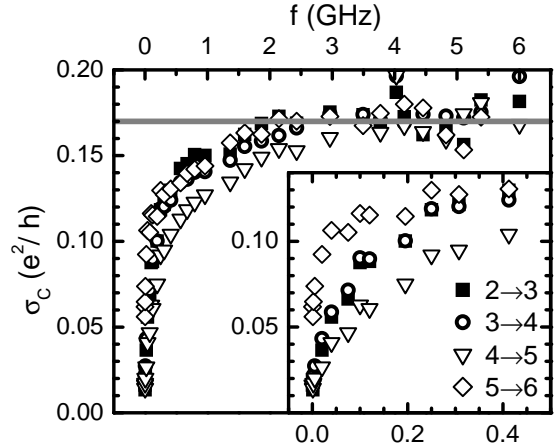}}}
  \end{center}
  \caption{Peak values of the real part of conductivity for different
    plateau transitions.}
  \label{amplitude}
\end{figure}
The reason for this disagreement lies in the sample geometry: The two
point Corbino geometry is sensitive to contact effects. An accumulation or
depletion zone along the ohmic contacts 
leads to edge modes not expected for ideal contacts. 
Such effects were observed e.g. in \cite{sachrajda96}, and
in recent work direct imaging of an edge
structure in a Corbino geometry was performed \cite{ahlswede00}.
This edge structure leads to an additional resistance in series with the 2DES
and therefore a reduced total DC-conductance of the sample. The DC-transport
mechanism from the contacts into the edge structure and further into the 
undisturbed 2DES is probably governed by tunneling processes. 
This explains
the low values of the critical conductivity, which also depend
on the filling factor,
measured in most experiments using Corbino geometry.

For AC-transport an additional transport mechanism from contact to 
undisturbed (bulk) 2DES is opened. This might be capacitive coupling as hinted
by $\im(\sigma)$ in an intermediate frequency range represented by 
$f=300$~MHz in figure \ref{data}. At sufficiently 
high frequencies the additional
edge series resistance becomes small compared to the resistance of the 
bulk 2DES and the conductance measurement yields the true conductivity of
the electron system. For our measurement this is true for $f>2$~GHz. 
Since we got rid of the disadvantages of Corbino geometry, namely the 
contact resistance, by applying high frequency, we are left with an advantage
in comparison to Hall geometry: Using Corbino geometry we have direct access 
to the longitudinal
conductivity, while for Hall geometries it is necessary to invert the
resistivity tensor with possible errors due to geometry and inhomogeneities.

For high frequencies $\sigma_c$ scatters
around $\sigma_c\approx 0.17 e^2/h$ and shows no further systematic frequency 
or filling factor dependence (figure \ref{amplitude}).
Also the temperature dependence of $\sigma_c$ at these
high frequencies is found to be negligible, whereas $\sigma_c$ shows
a strong decrease with decreasing temperature at low frequencies. Again this
behaviour can be modeled by a strongly temperature dependent edge resistance
in series with the intrinsic 2DES resistance.

At high frequencies
the critical conductivity $\sigma_c$ follows the predictions
of scaling theory,
but is still significantly lower than the proposed universal value 
$\sigma_c=0.5\,e^2/h$.
It is of the same order of magnitude as the
values found by Rokhinson \textsl{et al.} \cite{rokhinson95} in one of the
few experiments without apparent filling factor dependence. 
Believing in
universality one possible explanation for our experimental findings was
given by Ruzin, Cooper and Halperin \cite{ruzin96}. They showed that 
macroscopic inhomogeneities of the carrier density would lead to a
critical conductivity deviating from its universal microscopic value.

\begin{figure}[tb]
  \begin{center}
  \resizebox{\plotwidth}{!}{\rotatebox{270}{\includegraphics{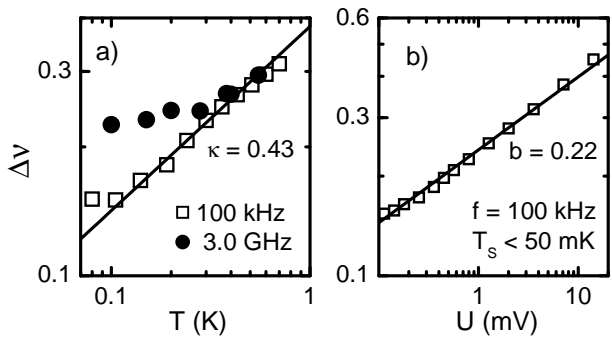}}}
  \end{center}
  \caption{Plateau transition width for transition
      $\nu = 3 \rightarrow 4$ defined as full width at half maximum (FWHM) 
      of the conductivity peak and 
      scaling analysis of a) temperature and 
      b) voltage dependence.}
  \label{temperature}
\end{figure}

%\mysection{Temperature scaling}
%
Leaving the critical point the next step is an ana\-ly\-sis of the 
transition width $\Delta \nu$ between quantum Hall plateaus. Before
heading to frequency scaling the first step is to test for
temperature scaling. Fi\-gure \ref{temperature}a) shows the temperature 
dependence of the plateau transition width $\Delta \nu$
plotted on logarithmic scale. The
plot is representative for all transitions in the filling factor 
range $\nu=2$ to $\nu=6$. The low frequency curves are well described
by a power law $\Delta \nu\propto T^\kappa$ with best fit results 
$0.39 < \kappa < 0.45$ for different transitions. This result was
tested at different frequencies (100~kHz and 300~MHz) and fits the 
commonly measured exponent $\kappa =p/2\gamma=0.43$ which is expected for 
$\gamma=2.3$  and $p=2$. 
Our data does not fit the linear dependence found in an experiment of Balaban
\textsl{et al.} \cite{balaban98}.
Using the scaling behaviour of the transition width as low
temperature thermometry for the electron system we estimate an electron
temperature $T_e$ for a base temperature $T_s < 50$~mK
between 100~mK and 150~mK, slightly dependent on magnetic field.

A second confirmation of scaling behaviour is given by the voltage 
dependence of $\Delta\nu$ shown in figure \ref{temperature}b):
It follows a power law $\Delta \nu \propto U^b$. Interpretation due to
a voltage dependent effective temperature $T_e \propto U^a$ with $a=b/\kappa$
leads to $a=2/(2+p)$.  An exponent $a=0.5$ 
equivalent to $p=2$ leads to $b=0.22$ which
fits our data.

The second data set in figure \ref{temperature}a) with $f=3$~GHz
represents the situation at high frequencies with $\Delta \nu$ defined
as the FWHM of the real conductivity $\re(\sigma)$: At low temperatures the 
transition width is no longer temperature dependent, but is determined by 
the frequency. 

\begin{figure}[tb]
  \begin{center}
  \resizebox{0.8\plotwidth}{!}{\rotatebox{0}{\includegraphics{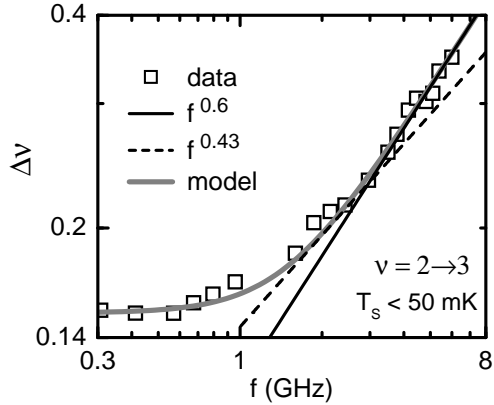}}}
  \end{center}
  \caption{Scaling analysis of the frequency dependence of the plateau 
    transition width.}
  \label{scaling}
\end{figure}

%\mysection{Frequency scaling}
%
In figure \ref{scaling} the frequency dependence of $\Delta \nu$ is plotted
for plateau transition $\nu=2\rightarrow 3$. As shown in the previous section
the frequency governs the transition width for $f\geq 3$~GHz while for 
$f \leq 1$~GHz the electron temperature leads to saturation. This restricts
a conventional scaling analysis to a frequency range $f=3-6$~GHz. A power
law fit $\Delta \nu \propto f^c$ in this range is shown as straight line 
and leads to an exponent $c=1/z\gamma=0.6\pm 0.1$, 
which is higher than the expected 0.43 for $z=1$ and $\gamma\approx 2.3$.
For comparison a power law with this exponent is plotted as dashed line. 
It is clearly less favourable than the higher exponent.
To overcome the unsatisfactory small fitting range it is necessary to 
incorporate both frequency and temperature into the scaling analysis.
We chose as an ansatz $L_{\text{eff}}^{-2} = L_\Phi^{-2}+L_f^{-2}$. 
The motivation
is an addition of scattering rates: The power law of the phase coherence
length $L_\Phi\propto T^{-p/2}$ follows from
$L_\Phi=\sqrt{D/\Gamma_\Phi}$ with the inelastic scattering rate
$\Gamma_\Phi$ and diffusion constant $D$. 
Interpreting in analogous way $\Gamma_f = D/L_f^2$ 
as frequency scattering rate
the Ma\-thies\-sen rule $\Gamma=\sum \Gamma_i$ leads to the proposed ansatz.
The resulting transition width is
\begin{equation*}
  \Delta \nu = \Delta \nu_0 \left(\left(\frac{T}{T_0}\right)^{p} +
    \left(\frac{f}{f_0}\right)^{2/z}\right)^{1/(2\gamma)}.
\end{equation*}
One of the parameters $\Delta\nu_0$, $T_0$ and $f_0$ can be chosen
arbitrarily. Here we choose $T_0=1$~K and use $\Delta\nu_0$, $p=2$ and
$\gamma=2.3$ from temperature and voltage scaling. A least square fit
with remaining parameters $f_0$ and $z$ for transitions in the filling 
factor range $\nu=2$ to $\nu=6$ leads to $z = 0.75\pm0.1$ compatible with
an exponent $c\approx 0.6$ for pure frequency scaling and deviating
from the expected dynamical exponent $z=1$. The fit is shown
as grey line in figure \ref{scaling}.

%\mysection{Summary}
%
In conclusion we were able to measure the frequency dependent 
complex conductivity of a 2DES
at quantizing magnetic fields and temperatures $T\leq 100$~mK up to 6~GHz.
This allows us to overcome contact effects and to
measure the critical conductivity at the quantum Hall plateau transition 
which deviates from the proposed universal value. Second we performed a
scaling analysis of the plateau transition width and incorporated a
two parameter ansatz, taking into account both temperature and frequency. 
We find significantly different scaling exponents for these two parameters
which are so far not understood theoretically.

\end{document}